\documentclass[sigconf, nonacm, amssymb]{acmart}

\usepackage{tikz}
\title[From Large Language Models to Databases and Back]{From Large Language Models to Databases and Back\\ {\em A discussion on research and education}}
\author[Sihem Amer-Yahia, Angela Bonifati, Lei Chen, Guoliang Li, Kyuseok Shim, Jianliang Xu, Xiaochun Yang]{Sihem Amer-Yahia (CNRS, Univ. Grenoble Alpes, France), Angela Bonifati (Univ. Lyon 1, CNRS Liris, France), Lei Chen (Hong Kong University of Science and Technology, China), Guoliang Li (Tsinghua University, China), Kyuseok Shim (Seoul National University, Korea), Jianliang Xu (Hong Kong Baptist University), Xiaochun Yang (Northeastern University, China)}
\date{April 2023}

\begin{document}

\begin{abstract}
This discussion was conducted at a recent panel at the 28th International Conference on Database Systems for Advanced Applications (DASFAA 2023), held April 17-20, 2023 in Tianjin, China.
The title of the panel was ``What does LLM (ChatGPT) Bring to Data Science Research and Education? Pros and Cons''. It was moderated by Lei Chen and Xiaochun Yang. The discussion raised several questions on how large language models (LLMs) and database research and education can help each other and the potential risks of  LLMs.  
\end{abstract}

\maketitle

\section{LLMs and Databases}

In recent years, large language models (LLMs) have garnered increasing attention from both academia and industry due to their potential to facilitate natural language processing (NLP)  and generate high-quality text. Despite their benefits, however, the use of LLMs is raising concerns about the reliability of knowledge extraction. The combination of database research and data science has advanced the state of the art in solving real-world problems, such as merchandise recommendation and hazard prevention. In this  discussion, we  explore the challenges and opportunities related to LLMs in database and data science research and education.

{\bf LLMs for database research.} 
LLMs have proven to be highly useful for language writing, as they can identify and correct grammar errors. Additionally, LLMs can serve as a valuable resource for knowledge acquisition and analysis. However, the accuracy of extracted knowledge is not guaranteed, and bias can  be introduced, leading to potential inaccuracies in the data analysis process.

LLMs can be highly effective in data preparation and labeling tasks, such as text mining, text parsing, keyword extraction, and sentiment analysis. It also has great potential to improve feature extraction, selection, and parameter tuning. 

{\bf Database research for LLMs.} 
Database research can support LLM development  from data cleaning and preprocessing to training and optimization. For instance, domain-specific knowledge can be incorporated into training data to create more accurate and reliable models. Furthermore, database research can be used to optimize prompt engineering to improve the effectiveness of LLM.

{\bf LLMs for education.} 
LLMs can be used to reform database education and address the challenges and concerns related to their use. LLMs can provide students with a wealth of knowledge and practical skills, such as techniques for handling dirty data. However, care must be taken to ensure that the information and programming styles provided by LLMs are accurate and free from bias or misinformation. As such, it is crucial to consider how to detect plagiarism when people use LLMs to generate scientific articles, papers, or assignments. We conclude that, although there are challenges associated with the use of LLMs in database research and education, these can be addressed through careful research and thoughtful integration of LLMs into the data science curriculum.

{\bf ChatGPT, today's famous LLM.} A Generative Pre-Trained Transformer (GPT) is a language model relying on deep learning that is designed to take a text-based input and generate a natural human-like text. Chat Generative Pre-trained Transformer (ChatGPT) is a chatbot released by OpenAI in November 2022 and is built on top of OpenAI's GPT-3.5 large language model (LLM). ChatGTP has been trained on a large body of text from a variety of sources in 2020 and can write any form of text such as essays, poems, paragraphs and computer programs.  It is able to understand and generate human-like natural language with a high level of accuracy and fluency. A new version based on GPT-4 was released on March 2023 and is available for paid subscribers on a limited basis.

While ChatGPT can help people with a lot of well-suited tasks such translation of foreign languages, summarization of a text and generation of human-like conversational responses, it has several drawbacks. Since large language models
perform the task of predicting the next word in a series of words by replicating common patterns of trained texts, it produces the output text without any concern of originality, plagiarism and privacy. Furthermore, since the training text data is derived from publicly available data before 2021, it cannot provide accurate information in a timely manner and  may not know the most up-to-date information. Moreover, ChatGPT does not seem to be yet able to perform well complicated mathematical calculations or high-level problem-solving tasks. Since its outputs may contain false or outdated information, we should carefully evaluate the outputs of ChatGPT and use them cautiously.

\section{Pros and Cons of LLMs for Research and Education}


\textbf{LLMs for Data Science Research.}
Calculators and word processors are useful tools for people that allow not to worry about complex arithmetic calculations and incorrect spellings/grammars as well as citation labels of their writing, respectively. The positive implications of using both tools are that people can focus and concentrate more on the content of their work without worrying about inaccurate calculations or spelling/grammatical errors. 

Similarly, data scientists can utilize ChatGPT for their data science research to focus more on high-level creative thinking including getting the big picture, original idea generation, analytical thinking and problem solving. For example,
they can utilize it to summarize the texts about related works, learn about a particular research topic, brainstorm about research directions for their new project and improve their writing skills of technical papers. ChatGPT can also assist non-native data scientists in understanding, interpreting and writing English texts. 
On the other hand, they can use ChatGPT to produce a high-quality code with explanations and improve their coding skills. 
It can even help a data scientist rewrite his old code in a programming language to an equivalent code in another programming language. Note that an important skill required for data scientists is the ability to produce a good quality of code. Thus, instead of spending time in learning how to code or producing code for data analysis, data scientists can concentrating more on their research by utilizing ChatGPT.


While more training data is likely to produce a more accurate model \cite{banko-brill-2001-scaling}, there are many applications such as named entity recognition \cite{wang2021meta}, relation extraction \cite{jung-shim-2020-dual} and image classification \cite{NEURIPS2021_995693c1} where producing a large-scale training data by manual labeling is expensive and time-consuming. To quickly obtain a large-scale training data with low cost, one approach is to use weak supervision that automatically annotates unlabeled data by heuristic rules or machine learning models. For example, one of the most popular techniques for weak supervision is distant supervision that utilizes external knowledge bases to produce weak labels \cite{DBLP:conf/kdd/LiangYJEWZZ20}. We can also utilize ChatGPT as an alternative method for weak supervision. For instance, ChatGPT was recently investigated to augment training data for few-shot classification by rephrasing each sentence in the training data into multiple similar sentences \cite{dai2023auggpt}. 

{\bf Data Science Research for LLMs.} Since an LLM model is only as knowledgeable as the training texts that have been provided for learning, its knowledge is limited according to the training data and it may become unfair by absorbing the biases from the training data. Furthermore, it lacks the capability of ethical thinking too. Thus, developing learning techniques to overcome such handicaps of LLM models will be very helpful to LLMs. Publicly available text data on the Web has a lot of sensitive information and training LLMs with public data can thus disclose sensitive and private information of people. On the other hand, as users input more data with conversations into ChatGPT, it may potentially leak the sensitive information to other users of ChatGPT. Thus, developing the privacy preserving schemes, such as the differential privacy, with high utility for training LLM modles will help LLMs to protect the privacy of individuals. 



{\bf LLMs for Computer Science and Data Science Education.} ChatGPT can be a useful tool for both disciplines and we need to reform the curricular to include ChatGPT. In \cite{cs-edu}, opportunities of utilizing ChatGPT are addressed and several ChatGPT-based tasks are suggested for computer science education.  For instance, teachers can ask students to generate a code for a given problem, and then explain, analyze and improve the code. In addition, we can also ask students to write their own code for the same problem and find the similarities as well as differences between two codes. 

Students can utilize ChatGPT to summarize/understand/learn the texts about existing works for a particular research topic,  enhance their coding as well as debugging skills to generate a high quality code,  brainstorm about research topics and improve their writing skills of technical papers. Thus, students can utilize ChatGPT to focus more on high-level creative thinking including getting the big picture, original idea generation, analytical thinking and the detailed steps of their methods to solve a given problem. To do so, since the outputs of ChatGPT may contain false or outdated information, students should learn how to use ChatGPT effectively and cautiously. 

While there are many advantages of including ChatGPT in the curriculum, students who consistently depend on ChatGPT may lose or cannot improve their skills of summarizing the texts, searching for relevant materials about a particular topic and writing technical papers by themselves. Furthermore, while ChatGPT is proficient in generating fluent text, it may produce the contents with lack of clarity as well as originality. Moreover, the outputs of ChatGPT may contain false or outdated information, and may even present a plagiarized writing from another source without citing properly. Thus, we need to provide precise guidelines of using ChatGPT to students so that they can learn how to use ChatGPT effectively and utilize the ChatGPT outputs with caution. 


{\bf Is ChatGPT charming?} Have you ever watched the movie ``Cyrano''? It is a story about a man who sent love letters to a woman that were actually written by another man with a good skill of writing romantic love letters to a woman. When you start chatting with someone for the first time on an online dating site or dating app, if you feel too attracted to the person, watch out - you may actually be talking with ChatGPT!

\section{What Can and Cannot LLM Do for Databases?}

As we all know, LLMs typically report probabilistic results but cannot be used to report fully deterministic results~\cite{natureprob,mitprob}. Therefore, LLM can be leveraged to handle inexact data/query processing problems that can tolerate approximate results, e.g., approximate query processing and data integration. However, it is hard to use LLMs to support exact data/query processing components, e.g., query answering and query rewriting. In the following, we discuss how to use LLMs to support exact and inexact data/query processing.

{\bf LLMs for Database Research.} For most of database problems, e.g., data discovery, data cleansing, and data integration, the users are satisfied with approximate results. The optimization goal is to improve the generalizability, efficiency and quality. Intuitively, we can utilize LLMs to improve the generalizability. However, several challenges arise. The first challenge is automatic prompt engineering that automatically generates appropriate prompts to guide LLMs to find correct answers. LLMs have limitations on the number of token constraints and long latency, and the automatic prompt engineering tool should optimize these two factors. The second challenge is how to integrate domain knowledge into LLMs.  Current LLMs use  open Web corpus to pretrain a large language model and thus can well support data cleaning and integration on Web data but cannot effectively support vertical domains that are absent on open Web. Hence, the challenge is to fine-tune LLMs to support domain knowledge or use prompt engineering to teach LLMs to do this. The third challenge is how to combine LLMs and existing data science tools, as it is expensive to call LLMs and it is beneficial to utilize some existing tools to reduce the cost. For example, there are many good database tools, e.g., blocking tools and entity-matching tools, and we can design tool learning that enable LLMs to call effective tools to reduce the cost. 

{\bf Database Research for LLMs.} The theory and model architecture of LLMs are almost mature, and the researchers and scientists that are working on LLMs focus on providing high-quality data to train LLMs, e.g., discovering data, cleaning data and integrating data. There exists a plethora of tools for the above database tasks that can be used to prepare the data on which LLMs are trained. 
A challenge is how to make a win-win loop between database systems and LLMs, which uses database techniques to provide high-quality of LLMs and uses LLMs to optimize the database tools.


\textbf{Querying data with natural language.} 
An interesting application is Text-to-SQL, which converts natural language queries into SQL statements. This has been a long-studied research problem. The state-of-the-art is currently a work presented at AAAI 2023~\cite{li2023resdsql}, which achieved an accuracy of 79.9\% using a seq2seq pre-trained language model. With  continuous prompts, 
ChatGPT enables users to interactively refine the generated SQL queries and could further improve their accuracy. This unique feature presents a potential opportunity to integrate LLMs with existing Text-to-SQL techniques to generate more precise SQL statements.

Additionally, ChatGPT allows users to query a dataset with natural language. 
This could  eliminate the need for SQL queries and increase the efficiency of data retrieval and analysis for certain applications, which poses the question of whether SQL remains necessary or if it is possible to translate text into a query evaluation plan for database result evaluation. 
The integration of LLMs with database techniques has the potential to open up new opportunities for research and advancement in the field of data science.

{\bf LLMs for Logical Query Optimization.}  Logical query optimization (e.g., query rewriting) aims to translate a query plan to an optimized query plan (possibly with a lower cost but without guarantee). It seems that LLMs can be used to support this problem. But the key challenge is that the translated query plan should be exactly equivalent to the original plan. Therefore, there are several possible solutions. The first uses LLMs to obtain an optimized query plan and then verifies the equivalence using existing techniques (and then keeps the equivalent query and drops the in-equivalent one). The second uses LLMs to optimize the using of query rewriting rules, including discovering new query-rewrite rules and judiciously using the rules (including whether to use a rule and to determine the order of using different rules). 


{\bf LLMs for Physical Query Optimization.}  Different from logical query optimization, physical query optimization should utilize the physical database statistics and it is vital to provide these information to LLMs. However, the current LLMs cannot effectively support numerical values. Two challenges arise in this context. The first is to fine-tune the LLMs that enables LLMs to support numerical statistics. The second is to embed database statistics into the prompt to facilitate that LLMs can use such information to get an optimized physical plan.  There are also some other similar problems that should utilize physical statistics, e.g., knob tuning, index/view advisor, and query diagnosis. 

{\bf Database Tuning for LLMs.}
Database knob tuning\cite{ituned} is important to achieve high performance of database systems. Traditionally, database administrators (DBAs) tune database systems.  Since the number of knobs in database systems increases as database systems become more sophisticated, it is difficult for DBAs to tune the database systems by considering all possible values of knobs \cite{ottertune}.  To overcome the drawbacks, auto-tuning methods were proposed to find an optimal configuration of database systems without human intervention \cite{ituned, ottertune}. Database knob tuning techniques developed for database systems can be useful to find an optimal configuration of an LLM for applications. 

{\bf LLMs for Database Storage and Transactions.} Both database storage and transactions are deterministic and we are pessimistic that LLMs cannot be used to optimize them.

\section{LLMs (badly) need integrated data and reasoning}

{\bf LLMs and data integration} To understand the differences between LLMs and databases, let us compare them with the process of integrating heterogeneous data sources. Data integration is a long-standing research problem in data management  \cite{BBR11,StonebrakerI18}. Schema mapping, data deduplication, schema and data fusion are all tasks that involve considering up-to-date data sources, as well as personal and proprietary data. By leveraging the inherent semantics of mappings (correspondences between queries or views on different data sources), these tasks do not need re-training on large corpuses of data and can easily capture new incoming data. As recently argued in a vision paper, these data of different nature are not considered so far in the LLM \cite{abs-2304-04576}. Moreover, integrated data can easily cater for privacy constraints and become trustworthy, for instance by blending mappings with policy views \cite{BonifatiCT21} or by having humans as first-class citizens in the data integration process \cite{ChiticariuT06,BonifatiCCT17,AriouaB18}. 

Provenance and lineage information, in particular why, how and where provenance characterizing query results \cite{CheneyCT09}, could serve the need of filling an existing gap of large language models, missing the key capability of locating the sources of information.  

But what is missing in LLMs to take advantage of databases and integrated data sources? The following is a non-exhaustive list of 
missing features (and the reader is invited to add more): 

\begin{description}
\item[-] Need to retain provenance and schema information as well as other kinds of metadata, which is not merely raw data and should not be unified with raw data; 
\item[-] Need to process and compute provenance throughout the learning process and be able to annotate the results with provenance information (and, thus, citation sources); 
\item[-] Need to capture privacy constraints and privacy policies in the data acquisition and data fusion process. 
\end{description}


{\bf LLMs and Graphs. } Graphs are a great source of knowledge, which is typically curated by humans and, as such, can be seen as high-quality and trustworthy integrated information. Examples of such highly curated graphs are Wikidata and DBPedia, that are typically 
used by search engines \cite{Weikum21}, whereas LLM are trained on large text corpora, such as Wikipedia, books, news and open datasets. 
Knowledge graphs such as DBPedia and Wikidata can be navigated and queried by leveraging query endpoints. Queries collected at the endpoints allow to understand what the users search within the knowledge graphs and thus indirectly to characterize the underlying structure of the data. 

To illustrate the difference between semantics in graph databases and language models, we choose to confront a query from the DBPedia graph query logs with question answering in ChatGPT.

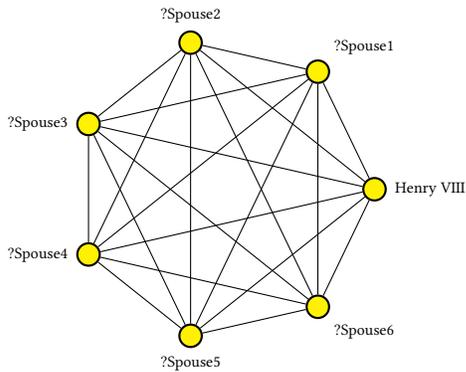
\begin{figure}[h]
\centering
\begin{tikzpicture}
  \tikzstyle{sommet}=[circle,draw,thick,fill=yellow]
        
	\node[sommet][label=0:{\scriptsize Henry VIII}] (h)  at (2,0) {};
	\node[sommet][label=51:{\scriptsize ?Spouse1}]   (s1) at (1.247,1.564) {};
	\node[sommet][label=90:{\scriptsize ?Spouse2}]   (s2) at (-0.445,1.95) {};
	\node[sommet][label=180:{\scriptsize ?Spouse3}]   (s3) at (-1.802,0.868) {};
	\node[sommet][label=180:{\scriptsize ?Spouse4}]   (s4) at (-1.802,-0.868) {};
	\node[sommet][label=-90:{\scriptsize ?Spouse5}]   (s5) at (-0.445,-1.95) {};
	\node[sommet][label=-51:{\scriptsize ?Spouse6}]   (s6) at (1.247,-1.564) {};
  
	\path[-]
  (h) edge (s1)
  (h) edge (s2)
  (h) edge (s3)
  (h) edge (s4)
  (h) edge (s5)
  (h) edge (s6)
  (s1) edge (s2)
  (s1) edge (s3)
  (s1) edge (s4)
  (s1) edge (s5)
  (s1) edge (s6)
  (s2) edge (s3)
  (s2) edge (s4)
  (s2) edge (s5)
  (s2) edge (s6)
  (s3) edge (s4)
  (s3) edge (s5)
  (s3) edge (s6)
  (s4) edge (s5)
  (s4) edge (s6)
  (s5) edge (s6)
;
\end{tikzpicture}
\caption{The Henry VIII query, a 7-clique containing one constant and six
  variables. All edges between Henry VIII and the variables are labeled
  ``dbpedia-owl:spouse'' and all edges between variables are labeled with the
  property path ``!dbpedia-owl:sameAs'' \cite{BonifatiMT20}.}\label{fig:henry}
\end{figure}

Figure \ref{fig:henry} shows the shape of a 7-node clique query involving Henry The VIII to his 6 wifes where all edges between Henry The VII and his wifes are labeled as ``dbpedia-owl:spouse'' and the edges between each pair of wifes are labeled with a property path \footnote{A property path is a path with a regular expression allowing to navigate graph data.}, namely ``!dbpedia-owl:same-As'' (to retrieve his distinct wifes).

\begin{figure}[h!]
    \centering
    \includegraphics[width=\linewidth]{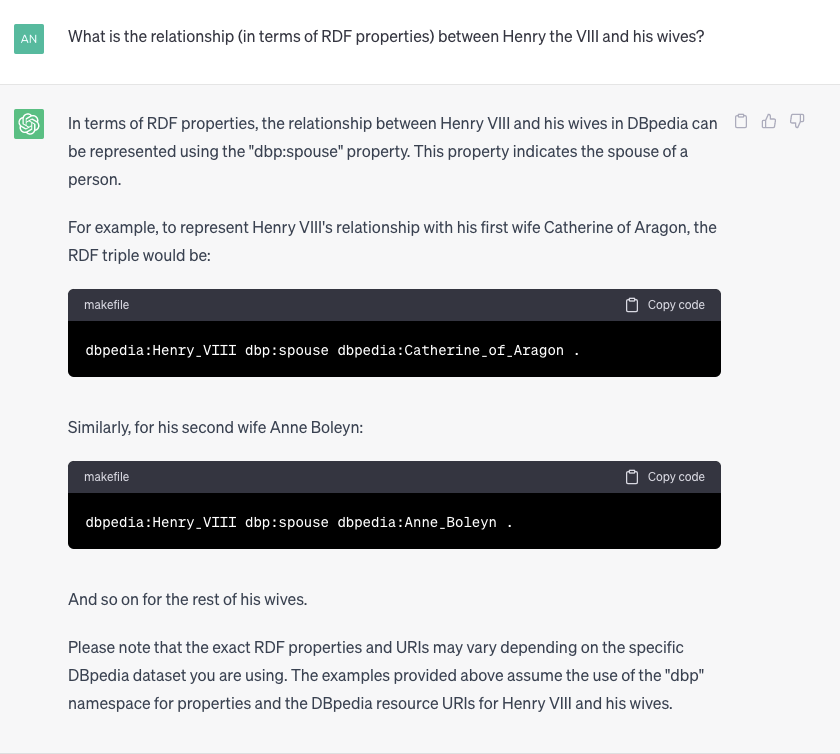}
    \caption{
    Snapshot of question answering on Henry The VIII's wives on ChatGPT.}
    \label{fig:HenryChat}
\end{figure}

Figure \ref{fig:HenryChat} shows question answering on ChatGPT on the possible relationships (in terms of RDF properties) between Henry The VIII and his 6 wives. 
It shows the RDF syntax of the ``spouse'' relationship but it does substantiate the result with the information about the source from which this information has been retrieved. ChatGPT cannot access databases such as DBPedia or Wikidata at the time being and, as text-based AI model is unable to draw diagrams on query shapes such as the one in Figure \ref{fig:henry}.  

{\bf LLMs and Reasoning.} Contrarily to LLMs, graphs enable symbolic reasoning, when logic-based existential rules and path queries are used to augment existing knowledge graphs with additional inferred data \cite{CarralDKL19}. Moreover, database queries including graph queries return certain answers \cite{2018BFVY}, whereas output of LLM is highly uncertain depending on the frequency of values in the training data.  

Combining results from LLMs with graphs could improve the results of LLMs by unifying machine reasoning with symbolic and logic-based reasoning \cite{SakrBVIAAAABBDV21}.
The knowledge graph lifecycle with maintenance operations, update propagation and error fixing is nonexistent for LLMs. 
On the other hand, the outputs of LLM could be used to enrich knowledge graphs for question answering tasks by leveraging graph attention networks \cite{YasunagaRBLL21}.

 Concluding, we cannot disagree with Gary Markus' blog `Hoping for the Best as AI Evolves' \cite{10.1145/3583078}: ``Large language models lack mechanisms for verifying truth; we need to find new ways to integrate them with the tools of classical AI, such as databases, Webs of knowledge, and reasoning.".

\begin{figure*}[htpb]
    \centering
    \includegraphics[width=\linewidth]{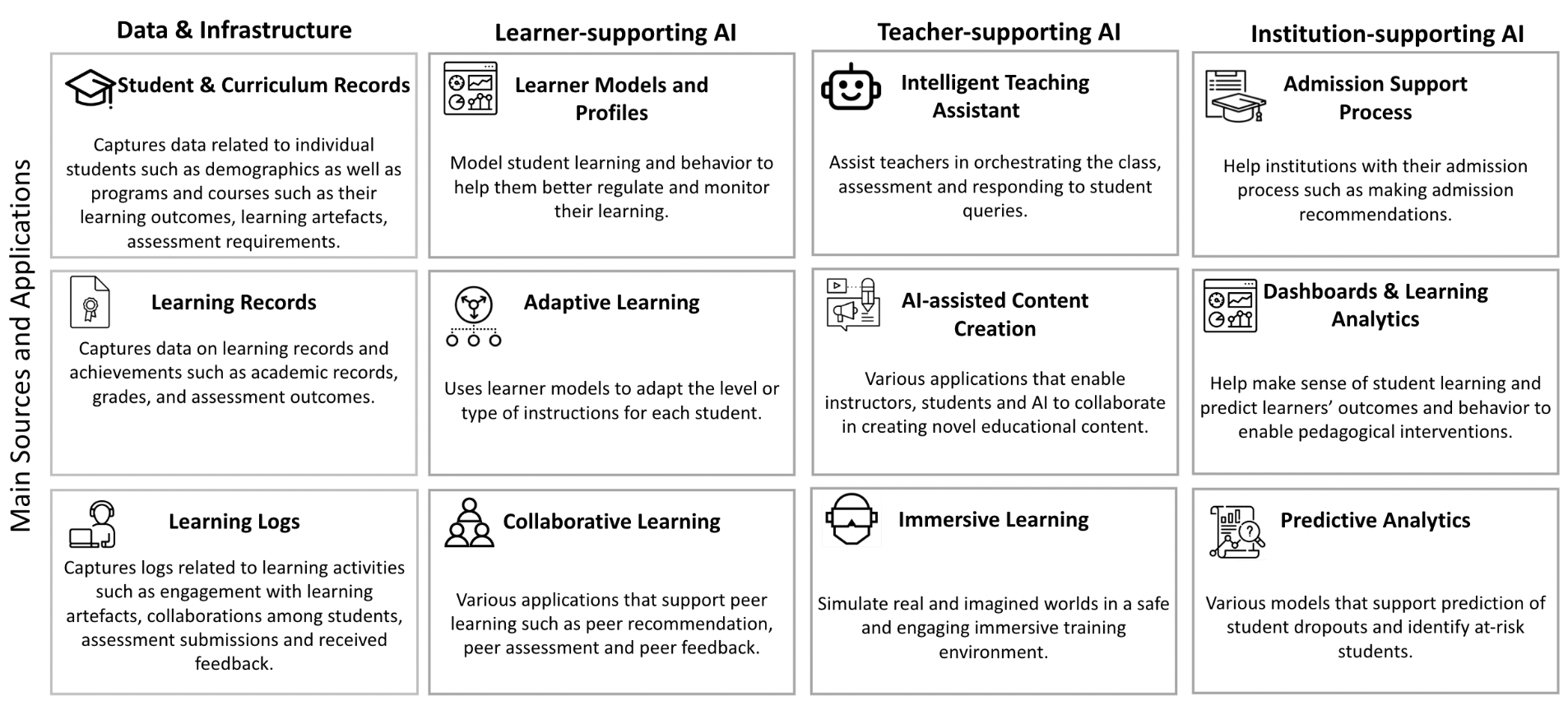}
    \caption{
    Overview of AIED research, focusing on the examination of underlying data and  applications~\cite{Sihem23}.}
    \label{fig:aied}
\end{figure*}

\section{LLMs as a research assistant}

\textbf{Helping with Scientific Writing.} LLMs like ChatGPT can help with scientific writing in several ways, such as proofreading, rewriting, summarization, and even suggesting titles for research papers. It can also help improve the language to better communicate research ideas and results, so as to facilitate research collaborations. However, the key challenge lies in creating effective prompts that generate high-quality responses. Below are some examples:
\begin{itemize}
\item 
\textit{Revise the following paragraph from the introduction of a Computer Science academic paper so the citations are kept and the text has a clear sentence structure.}
\item \textit{Here is the abstract of a paper. Suggest five creative titles.}
\item \textit{Write a 1-page sensational press release for this research.}
\end{itemize}

With well-crafted prompts, ChatGPT can deliver results that rival those of paid editing services.

\textbf{Assisting with Data Analytics.}  LLMs can  assist with data creation. It can generate data based on  input parameters, which can be useful in situations where  large amounts of synthetic data are needed for research. Additionally, LLMs can  help generate code for data analytics tasks~\cite{datacamp}. One such task is exploratory data analysis (EDA). For example, suppose we have a loan dataset that we want to perform EDA on. We can simply prompt ChatGPT with a request to write Python code to load and perform EDA on the loan dataset, and it will provide us with a code that we can use to analyze and visualize the data. 
ChatGPT can also assist with other data analytics tasks such as data cleaning and preprocessing, feature engineering, hyperparameter tuning, and model selection and evaluation. It can help save time and effort by automating some of these tedious tasks and focus on more complex aspects of data analytics.



\textbf{Limitation of Hallucination in LLMs.} Despite their many benefits, LLMs do have some limitations that need to be considered.
One of these limitations is their potential to generate incorrect content that appears plausible, known as ``hallucination''. This is particularly relevant in research paper writing, where ChatGPT may suggest non-existing references or provide inaccurate information. For example, when asked to recommend a paper on the topic of data science authored by Lei Chen, ChatGPT suggested a paper titled ``Crowdsourced Data Management: A Survey'' authored by Lei Chen, Reynold Cheng, Silviu Maniu, and Wang-Chien Lee and published in IEEE Transactions on Knowledge and Data Engineering, vol. 25, no. 9, pp. 1959-1977, September 2013. However, although this paper title does exist, it was not authored by Lei, rather by Guoliang et al. and published in 2016~(see~\cite{guoliang16}). In fact, Lei, Sihem, and Anand did author a survey paper on a related topic but their title is different (see~\cite{lei16}). To address this issue, one possible solution is to integrate ChatGPT with a knowledge graph and ground truth facts, which can help verify the accuracy of the generated information by cross-referencing it with existing data. 

\textbf{Incorporating  External Data into LLMs.}
While LLMs have access to a vast amount of data, their knowledge is still limited by the data they have been trained on. 
There are some recent efforts to incorporate external data into LLMs through the use of prompts. However, LLMs may have a limitation on the length of input they can process, which can impact their ability to understand complex or lengthy inputs. For example, the GPT-4 base version allows up to 8,192 tokens, which may not be sufficient for processing longer texts. One approach to overcome this limitation, known as chunking, is to split the longer text into smaller segments and process each segment separately~\cite{langchain}. It can be combined with traditional data science techniques such as embedding and indexing to improve the model's performance on longer inputs. Yet, chunking can also introduce challenges such as maintaining coherence and consistency between segments, which may require further research.

\textbf{Ethical and legal issues.} Ethical concerns such as bias, plagiarism, and data privacy and security are also significant issues when using LLMs. LLMs may be biased towards the data they were trained on, which can lead to unfair or inaccurate results. Additionally, there is a risk of privacy breaches if the input contains sensitive information.
Moreover, legal and copyright issues must also be considered when using LLMs. If the model generates copyrighted material without proper licensing, it could lead to legal repercussions. Thus, 
it is essential to responsibly and ethically use LLMs by thoroughly vetting and verifying generated content before using or publishing it. The new ACM policy requires disclosure of the use of generative AI tools for content generation in published work, with specific details regarding their usage provided in the acknowledgments section or elsewhere in the work~\cite{acm-policy}.

\section{LLMs and EDUCATION}
One area that is receiving both scientific and media coverage these days is the impact of LLMs on education. Several concerns have been raised about students using ChatGPT to complete tests, ChatGPT passing bar exams and professors using it to devise quizzes. As scientists, we focus on discussing LLMs in teaching and LLMs for doing research on education, that we refer to as LLM4ED.

{\bf Teaching LLMs.} First of all, we need to 
teach LLMs just like other models. This will contribute to demystifying them  and to raising awareness about their lack of transparency as well as their benefits and pitfalls. We also need to encourage our students to treat LLMs just like other recommendation engines. Their prediction accuracy should be tested keeping in mind that the best paper award at RecSys in 2019 showed that KNN outperformed 6 Deep Learning recommendation methods on MovieLens data~\cite{DBLP:journals/corr/abs-1907-06902}. These includes Collaborative Variational Autoencoder and and Neural Collaborative Filtering methods. Students need to learn to build on top of LLMs and treat them just like other models. In recent work, we built a meta-recommender that learns the best algorithm to apply given a (user,dataset) or a (user,question) pair~\cite{DBLP:conf/bigdataconf/BouarourBA21}. This approach could integrate LLMs as a recommendation option.

{\bf LLM4ED.} There are many challenges and opportunities of database research in education~\cite{DBLP:journals/pvldb/Amer-Yahia22, Sihem23}. An LLM could be modeled as a learner or to support learners, teachers or administrators. 

{\bf LLMs as Learners.} This would require to model learners' data and behavior. Figure~\ref{fig:aied} represents data about learners and learning artefacts that can be found in most education systems. This data is highly diverse. {\em Student and curriculum records} capture individual learners' records such as their demographics which are usually provided by learners at registration time as well as information on learning material such as artefacts, and assessment and outcome requirements. {\em Learning records} capture data on learners' achievements such as grades and assessment outcomes. {\em Learning logs} record learners' engagement with artefacts, feedback to learners and collaboration. This would encourage students to see LLMs as a peer from which to learn and to criticize.

{\bf LLMs for Learners, Teachers and Administrators.} This opens new opportunities for LLM-in-the-loop research in education. Figure~\ref{fig:aied} summarizes some research directions. For instance, in the context of collaborative learning, team formation algorithms could be revisited to consider LLMs as team members. In the context of developing an intelligent teaching assistant, LLMs are already in use for grading students which raises the question of accuracy and fairness. They are also used help teachers create content. In both grading and content creation, ensuring teachers' agency will allow them to guide the process and override automatic decisions.  LLMs can also be used for admission support and student dropout prediction analytics. A particular point of attention in all these applications is the study of bias in  ranking (student ranking) and in classification (student dropout prediction). 
Adding provenance to LLMs to better identify their sources of flaws would also address some of these concerns.  

\section{Conclusion}

Databases and LLMS are on either side of the spectrum of data science research.
Databases are collections of data, while LLMS are viewed as summaries and profiles of experiences based on data.  
Databases can help data scientists to query and statistically analyze the stored data accurately and efficiently. LLMs, on the other hand, are learned from textual data and can help data scientists to solve semantic application problems related to natural language. However, LLMs cannot guarantee an accurate or complete answer. 

We discussed the advantages and limitations that LLMs brought to data science and database research and education. 
Regarding the discussion of LLMs, the optimistic view is LLMs can facilitate most of the data science and data management tasks, including data discovery, data cleaning, data integration, and data visualization, and LLMs can bring great benefit to education. While the pessimistic view is LLMs are hard for data modeling, data analytics, and data interpretation, meanwhile, they might weaken learning skills and training LLMs could bring bias, plagiarism, privacy, legal, and copyright issues. People should be careful about the results obtained from an AI model and take them with a grain of salt.

We also showed our insights that data science and database research could also help LLMs, including how to use provenance and lineage information to fill the existing gap of LLMs, how to take advantage of databases and integrated data sources, how to unify machine reasoning with symbolic and logic-based reasoning by combining results from LLM with graphs, and how to combine model-centric and data-centric approaches altogether. 


\bibliographystyle{plain}
\bibliography{refs.bib}

\end{document}